\documentclass[12pt]{article}
\usepackage{amssymb,epsfig}

\renewcommand{\thefootnote}{\fnsymbol{footnote}}

\begin{document}

\title{
\begin{flushright}
\ \\*[-80pt] 
\begin{minipage}{0.2\linewidth}
\normalsize
%hep-th/yymmnnn \\
YITP-08-48 \\
KUNS-2145 \\*[50pt]
\end{minipage}
\end{flushright}
{\Large \bf 
Magnetized orbifold models
\\*[20pt]}}

\author{Hiroyuki~Abe$^{1,}$\footnote{
E-mail address: abe@yukawa.kyoto-u.ac.jp}, \ 
Tatsuo~Kobayashi$^{2,}$\footnote{
E-mail address: kobayash@gauge.scphys.kyoto-u.ac.jp} \ and \ 
Hiroshi~Ohki$^{3,}$\footnote{E-mail address: ohki@scphys.kyoto-u.ac.jp
}\\*[20pt]
$^1${\it \normalsize 
Yukawa Institute for Theoretical Physics, Kyoto University, 
% } \\ {\it \normalsize 
Kyoto 606-8502, Japan} \\
$^2${\it \normalsize 
Department of Physics, Kyoto University, 
Kyoto 606-8502, Japan} \\
$^3${\it \normalsize 
Department of Physics, Kyoto University, 
Kyoto 606-8501, Japan} \\*[50pt]}

\date{
\centerline{\small \bf Abstract}
\begin{minipage}{0.9\linewidth}
\medskip 
\medskip 
\small
We study (4+2n)-dimensional N=1 super Yang-Mills theory on 
the orbifold background with non-vanishing magnetic fluxes.
In particular, we study zero-modes of spinor fields.
The flavor structure of our models is different from one in  
magnetized torus models, and would be interesting in 
realistic model building. 
\end{minipage}
}

\begin{titlepage}
\maketitle
\thispagestyle{empty}
\end{titlepage}

%\tableofcontents

\renewcommand{\thefootnote}{\arabic{footnote}}
\setcounter{footnote}{0}

\section{Introduction}

Extra dimensional field theories, in particular 
string-derived extra dimensional field theories, 
play an important role in particle physics as well as cosmology.
When we start with extra dimensional theories, 
how to realize chiral theory is one of important 
issues from the viewpoint of particle physics.
Introducing magnetic fluxes in extra dimensions is 
one way to realize chiral fermions in 
field theories and superstring theories~\cite{Manton:1981es,
Witten:1984dg,Bachas:1995ik,Berkooz:1996km,Blumenhagen:2000wh,
Angelantonj:2000hi,Cremades:2004wa,Troost:1999xn}.
In particular, magnetized D-brane models are T-duals of 
intersecting D-brane models 
and several interesting models have been constructed 
within the framework of intersecting D-brane
models~\cite{Berkooz:1996km,Blumenhagen:2000wh,Angelantonj:2000hi,
Aldazabal:2000dg,Blumenhagen:2000ea,Cvetic:2001tj}.\footnote{
See for a review \cite{Blumenhagen:2005mu} and references therein.}

The generation number in magnetized extra dimensional models 
is fixed by the magnetic flux in the same way that 
the generation number in intersecting D-brane models is 
fixed by the intersecting number.
Yukawa couplings as well as other couplings in 
four-dimensional effective field theory can be 
calculated in  magnetized extra dimensional models 
as the overlapping integral of wave functions in the extra dimensional 
space.
We would obtain hierarchically small Yukawa couplings 
when the overlap integral of wave functions is  suppressed, that is, 
zero-modes are quasi-localized far away from each other in extra dimensions.
On the other hand, we would obtain Yukawa couplings
of ${\cal O}(1)$ 
when the overlap integral is not suppressed. 
That is an interesting aspect for the purpose 
to realize hierarchical patterns of 
quark/lepton mass matrices.
Indeed, Yukawa couplings on magnetized torus have 
been computed in \cite{Cremades:2004wa} and 
it has been shown that the results are the same as 
those in intersecting D-brane models.
However, it is still a challenging issue to derive 
proper generation numbers and 
realistic quark/lepton masses and mixing angles 
in magnetized extra dimensional models as well as 
intersecting D-brane models.

In this paper, we study orbifold models with non-vanishing 
magnetic fluxes, in particular N=1 super Yang-Mills theory 
on such a background.
Orbifolding the extra dimensions is another way to 
derive chiral theories \cite{Dixon:1985jw}.
We will show that four-dimensional effective field theories 
on magnetized orbifolds 
have a rich structure and they lead to interesting aspects, 
which do not appear in magnetized torus models.
In particular, it will be found that a new type of 
flavor structures can appear.
We also show semi-realistic models on magnetized orbifolds.

Organization of the paper is as follows.
In section 2, we study (4+2n)-dimensional N=1 super Yang-Mills theory, 
whose extra dimensions are torus with non-vanishing magnetic fluxes. 
We study fermionic and bosonic fields on the magnetized torus 
and the flavor structure of four-dimensional effective field theories.
Most of section 2 is a review.
(See e.g. \cite{Cremades:2004wa}.)
In section 3, we study the same D-dimensional N=1 super Yang-Mills
theory, but the toroidal orbifold background geometry with 
non-vanishing magnetic fluxes.
We study wavefunctions of fermionic and bosonic fields in 
the compact space, and flavor structure.
We will show semi-realistic models.
Section 4 is devoted to conclusion and discussion.
In appendix, we give two examples of models.

\section{Magnetized torus models}

\subsection{Extra dimensional super Yang-Mills theory}

Let us consider $N=1$ super Yang-Mills theory in $D=4+2n$ dimensions.
Its Lagrangian density is given by 
\begin{equation}
\label{eq:SYM-L}
{\cal L} = - \frac{1}{4g^2}{\rm Tr}\left( F^{MN}F_{MN}  \right) 
+\frac{i}{2g^2}{\rm Tr}\left(  \bar \lambda \Gamma^M D_M \lambda \right),
\end{equation}
where $M,N=0,\cdots, (D-1)$.
Here, $\lambda$ denotes gaugino fields, $\Gamma^M$ is the 
gamma matrix for $D$ dimensions and 
the covariant derivative $D_M$ is given as 
\begin{equation}
D_M\lambda = \partial_M \lambda - i [A_M, \lambda],
\end{equation}
where $A_M$ is the vector field.
Furthermore, the field strength $F_{MN}$ is given by 
\begin{equation}
F_{MN} = \partial_M A_N - \partial_N A_M -i[A_M,A_N].
\end{equation}

We consider the torus  $(T^2)^n$ as the extra dimensional 
compact space, whose coordinates are denoted by $y_m$ 
$(m=4, \cdots, 2n+3)$, while the coordinates of 
four-dimensional uncompact space $R^{3,1}$
are denoted by $x_\mu$ $(\mu=0,\cdots, 3)$.
We use orthogonal coordinates and choose the torus metric 
such that $y_m$ is identified by $y_m+n_m$ with $n_m=$ integer.
The gaugino fields $\lambda$ and the vector fields $A_m$ 
corresponding to the compact directions are decomposed as 
\begin{eqnarray}
\label{eq:gaugino-decomp}
\lambda(x,y) &=& \sum_n \chi_n(x) \otimes \psi_n(y), \\
\label{eq:vector-decomp}
A_m(x,y) &=& \sum_n \varphi_{n,m}(x) \otimes \phi_{n,m}(y).
\end{eqnarray}

\subsection{$U(1)$ gauge theory on magnetized torus $T^2$}

First, let us consider $U(1)$ gauge theory on $T^2$ 
with the coordinates $(y_4,y_5)$.
We study the non-vanishing constant magnetic flux $F_{45} = 2\pi M$.
We use the following gauge,
\begin{equation}
\label{eq:gauge}
A_4=0, \qquad A_5 = 2\pi My_4.
\end{equation}
Then, their boundary conditions can be written as 
\begin{eqnarray}
\label{eq:BC-gauge}
A_m(y_4+1,y_5)&=&A_m(y_4,y_5)+\partial_m \chi_4, \qquad   
\chi_4 = 2 \pi M y_5, \nonumber \\
A_m(y_4,y_5+1)&=&A_m(y_4,y_5)+\partial_m \chi_5, \qquad 
\chi_5 =0.
\end{eqnarray}

Now, we study the spinor field $\psi(y)$ with the $U(1)$ charge $q=\pm 1$ 
on $T^2$, which corresponds to the compact part in the 
decomposition (\ref{eq:gaugino-decomp}).
The zero-mode satisfies the following equation,
\begin{equation}\label{eq:Dirac-T2}
\tilde \Gamma^m(\partial_m -iqA_m)\psi(y) = 0,
\end{equation}
for $m=4,5$, where $\tilde \Gamma^m$ corresponds to 
the gamma matrix for the two-dimensional torus $T^2$, e.g.
\begin{equation}
\tilde \Gamma^4 = \left(
\begin{array}{cc}
0 & 1 \\
1 & 0 
\end{array}
\right), \qquad 
\tilde \Gamma^5 = \left(
\begin{array}{cc}
0 & -i \\
i & 0 
\end{array}
\right),
\end{equation}
and $\psi(y)$ is the two component spinor,
\begin{equation}\label{eq:two-spinor}
\psi = \left(
\begin{array}{c}
\psi_+ \\ \psi_-
\end{array}
\right).
\end{equation}
Because of (\ref{eq:BC-gauge}), the spinor field satisfies 
the following boundary condition,
\begin{eqnarray}
\label{eq:bc-1}
\psi(y_4+1,y_5) &=& e^{iq\chi_4}\psi(y_4,y_5) = 
e^{2\pi i qMy_5}\psi(y_4,y_5), \\
\psi(y_4,y_5+1) &=& e^{iq\chi_5}\psi(y_4,y_5) = 
\psi(y_4,y_5).
\label{eq:bc-2}
\end{eqnarray}
The consistency for the contractible loop, i.e.
%\begin{equation}
$(y_4,y_5) \rightarrow (y_4+1,y_5) \rightarrow (y_4+1,y_5+1)$,  
%\end{equation}
requires $M=$ integer.
Because of the periodicity along $y_5$, 
$\psi_{\pm}$ can be written by 
\begin{equation}
\psi_{\pm}(y_4,y_5) = \sum_n c_{\pm,n}(y_4)e^{2\pi i ny_5}.
\end{equation}
Suppose that $qM >0$.
Then, the solution for the zero-mode equation of $\psi_+$ 
is given by 
\begin{equation}
c_{+,n}(y_4) = k_{+,n} e^{-\pi qMy_4^2+2\pi n y_4},
\end{equation}
where $k_{+,n}$ is a constant.
Furthermore the boundary condition requires 
\begin{equation}
k_{+,n} = a_n e^{-\pi n^2/(qM)},
\end{equation}
and $a_{n+qM}$ is equal to $a_n$, i.e. $a_{n+qM}=a_n$.
Thus, there are $|M|$ independent zero modes of $\psi_+$, 
which have normalizable wave functions,
\begin{equation}\label{eq:zero-mode-wf}
\Theta^j(y_4,y_5) = N_je^{-M\pi y_4^2}\vartheta \left[
\begin{array}{c}
j/M \\ 0
\end{array} \right]\left( M(y_4+iy_5), Mi\right), 
\end{equation}
for $j=0,\cdots, M-1$, where $N_j$ is a normalization constant and 
\begin{equation}
\vartheta \left[
\begin{array}{c}
j/M \\ 0
\end{array} \right]
\left( M(y_4+iy_5), Mi\right) = \sum_n e^{-M\pi (n+j/M)^2 
+2\pi (n+j/M)M(y_4+iy_5)}, 
\end{equation}
that is, the Jacobi theta-function.
We can introduce the complex structure modulus $\tau$ by replacing 
the above Jacobi theta-function as 
\begin{equation}
\vartheta \left[
\begin{array}{c}
j/M \\ 0
\end{array} \right]
\left( M(y_4+iy_5), Mi\right) \rightarrow 
\vartheta \left[
\begin{array}{c}
j/M \\ 0
\end{array} \right]
\left( M(y_4+\tau y_5), M \tau \right).
\end{equation}
Thus, zero-mode wave functions depend on only 
the complex structure modulus, but not the overall 
size of $T^2$.
Furthermore, 
there is the degree of freedom to shift $y_m \rightarrow y_m + d_m$ 
with constants $d_m$.
They correspond to constant Wilson lines.

On the other hand, the zero-modes equation for 
$\psi_-$ can be solved in a similar way, but their wave functions 
are unnormalizable.
Hence, we can derive chiral theory by introducing magnetic fluxes.
When $qM <0$, $\psi_-$ has $|M|$ independent zero modes 
with normalizable wave functions, while zero modes for 
$\psi_+$ have unnormalizable wave functions.
Bosonic fields are analyzed in a similar way.
(See e.g. \cite{Cremades:2004wa}.)

\subsection{$U(N)$ gauge theory on magnetized torus $T^2$}

Here, we study $U(N)$ gauge theory on $T^2$.
Let us consider the following form of (abelian) magnetic 
flux
\begin{equation}\label{eq:F45-UN}
F_{45} = 2 \pi \left(
\begin{array}{ccc}
M_1 {\bf 1}_{N_1\times N_1} & & 0 \\
 & \ddots & \\
0 & & M_n {\bf 1}_{N_n\times N_n}
\end{array}\right),
\end{equation}
where ${\bf 1}_{N_a \times N_a}$ denotes $(N_a \times N_a)$ 
identity matrix.
This abelian magnetic flux breaks the gauge group as 
$U(N) \rightarrow \prod_{a=1}^n U(N_a)$ with $N=\sum_a N_a$.
The rank is not reduced by the abelian magnetic flux.
When we consider non-abelian magnetic flux, 
i.e. the toron background \cite{'t Hooft:1979uj},  
the rank can be reduced.\footnote{
See e.g. \cite{Alfaro:2006is,vonGersdorff:2007uz} and references therein.}
However, here we restrict ourselves to the abelian flux.

Now, let us study gaugino fields on this background.
We focus on the block including only $U(N_a) \times U(N_b)$ 
and such a block has the following magnetic flux,
\begin{equation}\label{eq:F-ab-block}
F_{45} = 2 \pi \left(
\begin{array}{cc}
M_a {\bf 1}_{N_a\times N_a}  & 0 \\
0 & M_b {\bf 1}_{N_b\times N_b}
\end{array}\right).
\end{equation}
We use the same gauge as (\ref{eq:gauge}), i.e.
\begin{equation}
A_4 = 0, \qquad A_5 = F_{45}y_4.
\end{equation}

Similarly, the gaugino fields $\lambda$ in $R^{3,1}\times T^2$ are 
decomposed as 
\begin{equation}
\lambda(x,y) = \left(
\begin{array}{cc}
\lambda^{aa}(x,y)   & \lambda^{ab}(x,y)  \\
\lambda^{ba}(x,y) & \lambda^{bb}(x,y) 
\end{array}\right).
\end{equation}
Furthermore these gaugino fields are decomposed as 
(\ref{eq:gaugino-decomp}), 
\begin{eqnarray}
\lambda^{aa}(x,y)  &=& \sum_n \chi^{aa}_n(x) \otimes \psi^{aa}_n(y), 
\quad 
\lambda^{ab}(x,y) = \sum_n \chi^{ab}_n(x) \otimes \psi^{ab}_n(y), 
\nonumber \\
\lambda^{ba}(x,y)  &=& \sum_n \chi^{ba}_n(x) \otimes \psi^{ba}_n(y), 
\quad 
\lambda^{bb}(x,y) = \sum_n \chi^{bb}_n(x) \otimes \psi^{bb}_n(y). 
\end{eqnarray}
Each of $\psi^{aa}$, $\psi^{ab}$, $\psi^{ba}$ and $\psi^{bb}$ 
is a two-component spinor $(\psi_+,\psi_-)^T$.
Their zero-modes satisfy 
\begin{eqnarray}
\left(
\begin{array}{cc}\label{eq:Dirac-U(N)-T2-1}
\bar \partial \psi_+^{aa} & 
[\bar  \partial +2\pi (M_a-M_b) y_4] \psi^{ab}_+ \cr \cr
  [\bar \partial +2\pi (M_b-M_a)y_4] \psi^{ba}_+ & 
\bar \partial \psi_+^{bb}  \\
\end{array}
\right) &=& 0, \\
%\end{equation}
%
%\begin{eqnarray}
\left(
\begin{array}{cc}\label{eq:Dirac-U(N)-T2-2}
\partial \psi_-^{aa} & [\partial -2\pi (M_a-M_b)y_4] \psi^{ab}_-  \cr \cr
[\partial -2\pi (M_b-M_a)y_4] \psi^{ba}_- & \partial \psi_-^{bb}  \\
\end{array}
\right) &=& 0,
\end{eqnarray}
where $\bar \partial = \partial_4 +i \partial_5$ and 
$\partial = \partial_4  -i \partial_5$.

The zero-modes of $\psi^{aa}$ and $\psi^{bb}$ correspond to 
four-dimensional massless gauginos for the 
unbroken gauge group $U(N_a)\times U(N_b)$.
Dirac equations of $\psi^{aa}(y)$ and $\psi^{bb}(y)$ 
in (\ref{eq:Dirac-U(N)-T2-1}) and (\ref{eq:Dirac-U(N)-T2-2}) do not 
include any magnetic fluxes.
That is, both of $\psi_{\pm}$ have the same zero-modes as 
those on $T^2$ without magnetic fluxes.

Next, we study spinor fields, $\lambda^{ab}$ and $\lambda^{ba}$,  
which correspond to bi-fundamental matter fields, $(N_a,\bar N_b)$ 
and $(\bar N_a,N_b)$ for the unbroken gauge group  
$U(N_a)\times U(N_b)$.
When $M_a - M_b >0$, $\lambda^{ab}_+$ and $\lambda^{ba}_-$ 
have $(M_a-M_b)$ zero-modes with normalizable wave functions, 
i.e. $\Theta^j(y_4,y_5)$ for $j=0,\cdots, (M_a-M_b-1)$ as 
(\ref{eq:zero-mode-wf}), but 
zero-mode wave functions of $\lambda^{ab}_-$ and $\lambda^{ba}_+$ 
are unnormalizable.
On the other hand, when $M_a - M_b <0$, 
$\lambda^{ab}_-$ and $\lambda^{ba}_+$ have $(M_b - M_a)$ normalizable 
zero-modes.
Hence, we obtain chiral theory.
We have the degree of freedom of the constant shift 
$y_m \rightarrow y_m + d_m$.

Similarly, we can analyze bosonic fields $A_m$.
In general, introduction of non-vanishing magnetic fluxes on $T^2$ 
breaks supersymmetry completely.

\subsection{$U(N)$ gauge theory on $(T^2)^3$}

Here, we extend the previous analysis to 
$U(N)$ gauge theory on $(T^2)^3$.
We consider the magnetic background, where 
only $F_{45}, F_{67}$ and $F_{89}$ are non-vanishing, 
but the others of $F_{mn}$ are vanishing.
Furthermore, $F_{45}, F_{67}$ and  $F_{89}$ are 
given by 
\begin{eqnarray}\label{eq:6D-flux}
F_{45} &=& 2 \pi \left(
\begin{array}{ccc}
M^{(1)}_1 {\bf 1}_{N_1\times N_1} & & 0  \\
 & \ddots & \\
0 & & M^{(1)}_n {\bf 1}_{N_n\times N_n}
\end{array}\right),  \nonumber \\
F_{67} &=& 2 \pi\left(
\begin{array}{ccc}
M^{(2)}_1 {\bf 1}_{N_1\times N_1} & & 0 \\
 & \ddots & \\
0 & & M^{(2)}_n {\bf 1}_{N_n\times N_n}
\end{array}\right),  \\
F_{89} &=& 2 \pi \left(
\begin{array}{ccc}
M^{(3)}_1 {\bf 1}_{N_1\times N_1} & & 0 \\
 & \ddots & \\
0 & & M^{(3)}_n {\bf 1}_{N_n\times N_n}
\end{array}\right). \nonumber 
\end{eqnarray}
This background breaks the gauge group $U(N)$ as 
$U(N) \rightarrow \prod_{a=1}^n U(N_a)$ with $N=\sum_a N_a$.

We can study gaugino fields on this background as 
a simple extension of the previous section 2.3.
That is, we focus on the block including only $U(N_a) \times U(N_b)$ 
and such a block has the following magnetic flux as 
(\ref{eq:F-ab-block}), 
\begin{equation}\label{eq:6D-F-ab-block}
F_{2i+2,2i+3} = 2 \pi\left(
\begin{array}{cc}
M^{(i)}_a {\bf 1}_{N_a\times N_a}  & 0 \\
0 & M^{(i)}_b {\bf 1}_{N_b\times N_b}
\end{array}\right),
\end{equation}
and  we use the following gauge  
\begin{equation}\label{eq:6D-gauge}
A_{2i+2} = 0, \qquad A_{2i+3} = y_{2i+2}F_{2i+2, 2i+3},
\end{equation}
for $i=1,2,3$.
Then, we decompose the gaugino fields $\lambda(x,y)$ as 
(\ref{eq:gaugino-decomp}), i.e. the four-dimensional part $\chi(x)$ 
and the $i$-th $T^2$ part $\psi_{(i)}(y_{2i+2},y_{2i+3})$, 
whose zero modes satisfy 
\begin{eqnarray}
\left(
\begin{array}{cc}
\bar \partial_i \psi_{(i)+}^{aa} & [\bar  \partial_i +2\pi
(M^{(i)}_a-M^{(i)}_b)y_{2i+2} ] 
\psi^{ab}_{(i)+} \cr \cr
  [\bar \partial_i + 2\pi(M^{(i)}_b-M^{(i)}_a)y_{2i+2}] \psi^{ba}_{(i)+} 
& \bar \partial_i \psi_{(i)+}^{bb}  \\
\end{array}
\right) &=& 0, \nonumber \\
%\end{equation}
%
%\begin{eqnarray}
 & & \\
\left(
\begin{array}{cc}
\partial_i \psi_{(i)-}^{aa} & 
[\partial_i - 2\pi(M^{(i)}_a-M^{(i)}_b)y_{2i+2}] 
\psi^{ab}_{(i)-}  \cr \cr
[\partial_i -2 \pi(M^{(i)}_b-M^{(i)}_a)y_{2i+2}] \psi^{ba}_{(i)-} & 
\partial_i \psi_{(i)-}^{bb}  \\
\end{array}
\right) &=& 0, \nonumber
\end{eqnarray}
where $\bar \partial_i = \partial_{2i+2}  +i \partial_{2i+3}$ and 
$\partial_i = \partial_{2i+2} -i \partial_{2i+3}$.

The gaugino fields, $\psi^{aa}$ and $\psi^{bb}$, for 
the unbroken gauge symmetry have no effect from magnetic 
fluxes in their Dirac equations.
Hence, they have the same zero-modes as those on 
$(T^{2})^3$ without magnetic fluxes.
On the other hand, $\psi^{ab}$ and $\psi^{ba}$ correspond 
to bi-fundamental matter fields, 
$(N_a, \bar N_b)$ and $(\bar N_a,N_b)$. 
For the $i$-th $T^2$ with $M^{(i)}_a -M^{(i)}_b >0$,
$\psi^{ab}_{(i)+}$ and $\psi^{ba}_{(i)-}$ have 
$|M^{(i)}_a - M^{(i)}_b |$ normalizable zero-modes, 
while $\psi^{ab}_{(i)-}$ and $\psi^{ba}_{(i)+}$ have 
no normalizable zero-modes.
When $M^{(i)}_a -M^{(i)}_b <0$, 
$\psi^{ab}_{(i)-}$ and $\psi^{ba}_{(i)+}$ have 
$|M^{(i)}_a - M^{(i)}_b |$ normalizable zero-modes.
Then, the total number of bi-fundamental zero-modes 
is given by $\prod_{i=1}^3|M_a^{(i)}-M_b^{(i)}|$ and 
all of them have the same six-dimensional chirality 
${\rm sign} \left[ \prod_{i=1}^3(M_a^{(i)}-M_b^{(i)})\right]$.
Since the ten-dimensional chirality of gaugino fields is fixed, 
bi-fundamental zero-modes for either $(N_a,\bar N_b)$ or 
$(\bar N_a,N_b)$ appear with a fixed four-dimensional chirality.
To summarize, the total number of bi-fundamental zero-modes 
for $(N_a,\bar N_b)$ is equal to 
\begin{equation}
I_{ab}=\prod_{i=1}^3(M_a^{(i)}-M_b^{(i)}),
\end{equation}
and their wave functions are given by a product
of two-dimensional parts, i.e. 
\begin{equation}\label{eq:zero-mode-6D}
\Theta^{i_1,i_2,i_3}(y) = \Theta^{i_1}(y_4,y_5) 
\Theta^{i_2}(y_6,y_7) \Theta^{i_3}(y_8,y_9),
\end{equation}
for $i_1 =0,\cdots, (M_a^{(1)}-M_b^{(1)}-1)$, 
$i_2 =0,\cdots, (M_a^{(2)}-M_b^{(2)}-1)$ and 
$i_3 = 0,\cdots, (M_a^{(3)}-M_b^{(3)}-1)$.
For $I_{ab} <0$, this means 
that there appear $|I_{ab}|$ independent zero modes for 
$(\bar N_a,N_b)$.
It is also convenient to introduce the notation, 
$I^i_{ab} \equiv M_a^{(i)}-M_b^{(i)}$.

Similarly, we can analyze bosonic fields corresponding to $A_m$ 
for $m=4, \cdots, 9$.
For generic values of magnetic fluxes, supersymmetry 
is broken completely.
However, when they satisfy the following 
condition \cite{Troost:1999xn,Cremades:2004wa},
\begin{equation}\label{eq:SUSY-condition-0}
\sum_{i=1}^3\pm \frac{M_a^{(i)}-M_b^{(i)}}{{\cal A}^{(i)}} =0,
\end{equation}
for one combination of signs, 
where ${\cal A}^{(i)}$ denotes the area of the $i$-th torus,
there appear massless scalar modes as well as massive modes and 
four-dimensional N=1 supersymmetry remains unbroken at least in 
the $a-b$ sector.
When we consider ${\cal A}^{(i)}$ as free parameters, 
we can realize the above supersymmtric condition
(\ref{eq:SUSY-condition-0}) for 
most cases by choosing proper values of ${\cal A}^{(i)}$.
For the case with the universal area, 
${\cal A}^{(1)}={\cal A}^{(2)}={\cal A}^{(3)}$, the above condition 
(\ref{eq:SUSY-condition-0}) reduces to  
\begin{equation}\label{eq:SUSY-condition}
\sum_{i=1}^3\pm (M_a^{(i)}-M_b^{(i)}) =0.
\end{equation}
In addition to (\ref{eq:SUSY-condition-0}), 
when one of them is vanishing, i.e.
$(M_a^{(i)}-M_b^{(i)}) =0$ and 
\begin{equation}
\sum_{j \neq i} \pm \frac{M_a^{(j)}-M_b^{(j)}}{{\cal A}^{(j)}} 
=0,
\end{equation}
four-dimensional N=2 supersymmetry is unbroken.
In these supersymmetric models, 
zero-mode profiles of bosonic fields are the same as 
their superpartners, that is, zero-mode profiles 
of fermionic fields.

\subsection{$U(N_a)\times U(N_b)\times U(N_c)$ models 
with three families}

Here, we consider an illustrating model with the unbroken 
gauge group $U(N_a)\times U(N_b)\times U(N_c)$, 
which is derived from ten-dimensional $U(N)$ super Yang-Mills 
theory on $R^{3,1}\times (T^{2})^3$ with the magnetic fluxes 
as in the previous section.
We assume that the magnetic fluxes satisfy the supersymmetric 
condition (\ref{eq:SUSY-condition-0}) and 
massless scalar fields appear as partners of 
massless spinor fields with bi-fundamental representations.
%four-dimensional N=1 supersymmetry remains unbroken.
In addition to supersymmetric vector multiplets 
for the gauge group $U(N_a)\times U(N_b)\times U(N_c)$, 
the massless spectrum of this model includes 
three types of bi-fundamental matter fields, 
$(N_a,\bar N_b)$, $(N_b,\bar N_c)$ and $(N_c,\bar N_a)$.
This class of models include the 
$SU(4)\times SU(2)_L \times SU(2)_R$ Pati-Salam model for 
$N_a=4, N_b=2$ and $N_c=2$.\footnote{
$U(1)^3$ would be anomalous and their gauge bosons 
would become massive through the Green-Schwarz mechanism.}
In this case, bi-fundamental matter fields $(4,2,1)$, 
$(\bar 4,1,2)$ and $(1,2,2)$ under 
$SU(4)\times SU(2)_L \times SU(2)_R$ include 
left-handed quarks and leptons in $(4,2,1)$, 
the up- and down-sectors of right-handed quarks and 
right-handed charged leptons and neutrinos in $(\bar 4,1,2)$ and 
up- and down-sectors of Higgs fields in $(1,2,2)$.
Indeed, in intersecting D-brane models 
it is a convenient way %in intersecting D-brane models 
that first one  constructs  the 
supersymmetric Pati-Salam model and then breaks it 
to the minimal supersymmetric standard model (MSSM) in order 
to realize the MSSM-like models within the framework of 
intersecting D-brane models.
(See e.g.  \cite{Cvetic:2001tj,Blumenhagen:2002gw} and 
references therein.)\footnote{
See e.g. for the Pati-Salam model from heterotic orbifold models 
\cite{Kobayashi:2004ud}, where $SU(4)\times SU(2)_L \times SU(2)_R$ 
is broken to the standard gauge group by vacuum expectation values 
of scalar fields, $(4,1,2)$ and $(\bar 4,1,2)$, 
while in the intersecting D-brane models 
$SU(4)\times SU(2)_L \times SU(2)_R$ is broken by splitting 
D-branes, that is, vacuum expectation values of adjoint scalar 
fields.}
For the purpose to derive a semi-realistic model, 
we consider the realization of three families of 
$(N_a,\bar N_b)$ and $(\bar N_a,N_c)$ matter fields,
i.e. $I_{ab}=I_{ca}=3$.
Yukawa couplings among $(N_a,\bar N_b)$ and $(\bar N_a,N_c)$ 
matter fields and $(N_b,\bar N_c)$ Higgs fields in four-dimensional 
effective theory are given by the overlap integral of 
zero-mode wave functions (\ref{eq:zero-mode-6D}) 
in extra dimensions \cite{Green:1987mn}, 
\begin{equation}\label{eq:Y-overlap}
Y^{ijk} = g \int dy~\Theta^{i_1,i_2,i_3}(y) \cdot \Theta^{j_1,j_2,j_3}(y) 
\cdot \Theta^{k_1,k_2,k_3}(y),
\end{equation}
in the canonically normalized basis.

Now, let us study $U(N_a)\times U(N_b)\times U(N_c)$ models,
which lead to three 
families of $(N_a,\bar N_b)$ and $(\bar N_a,N_c)$, i.e. 
$I_{ab}=\prod_{i=1}^3 I^i_{ab}=3$ and 
$I_{ca}=\prod_{i=1}^3 I^i_{ca}=3$.
First, it is straightforward to show that we can not 
derive three-family models, which satisfy 
the condition (\ref{eq:SUSY-condition}), 
that is, it is difficult to realize supersymmetric 
three-family models from ten-dimensional super 
Yang-Mills theory with the universal area of tori.
Thus, we study models with the condition (\ref{eq:SUSY-condition-0}).

For generic model building with the condition
(\ref{eq:SUSY-condition-0}), 
we can construct three-family models.
Magnetic fluxes leading to three families are classified 
into two classes.
One class corresponds to the following magnetic fluxes,
\begin{equation}\label{eq:3-f-model-1}
(|I^{(1)}_{ab}|,|I^{(2)}_{ab}|,|I^{(3)}_{ab}|)=(3,1,1), 
\qquad 
(|I^{(1)}_{ca}|,|I^{(2)}_{ca}|,|I^{(3)}_{ca}|)=(1,3,1), 
\end{equation}
and their permutations, and the other 
corresponds to 
\begin{equation}\label{eq:3-f-model-2}
(|I^{(1)}_{ab}|,|I^{(2)}_{ab}|,|I^{(3)}_{ab}|)=(3,1,1), 
\qquad 
(|I^{(1)}_{ca}|,|I^{(2)}_{ca}|,|I^{(3)}_{ca}|)=(3,1,1), 
\end{equation}
and their permutations.
Hence, we can realize the restricted flavor structure.
Moreover, the number of Higgs fields are constrained because 
$|I^{(i)}_{bc}|=|\pm I^{(i)}_{ab} \pm I^{(i)}_{ca}|$.\footnote{
Similar results have been obtained in 
intersecting D-brane models \cite{Cremades:2003qj,Higaki:2005ie}, 
which are T-duals of magnetized D-brane models.}
For example, in the first class of models (\ref{eq:3-f-model-1}) 
we would obtain 
\begin{equation}
|I^{(1)}_{bc}| = 4~~{\rm or~~} 2, \qquad 
|I^{(2)}_{bc}| = 4~~{\rm or~~} 2, \qquad 
|I^{(3)}_{bc}| = 2~~{\rm or~~} 0,
\end{equation}
and the total Higgs number would be equal to 
$\prod_i |I^{(i)}_{bc}|=0,8,16,32$.
On the other hand, in the second class of models
(\ref{eq:3-f-model-2}), we would obtain 
 \begin{equation}
|I^{(1)}_{bc}| = 6~~{\rm or~~} 0, \qquad 
|I^{(2)}_{bc}| = 2~~{\rm or~~} 0, \qquad 
|I^{(3)}_{bc}| = 2~~{\rm or~~} 0,
\end{equation}
and the total Higgs number would be equal to 
$\prod_i |I^{(i)}_{bc}|=0,24$.
Thus, the total Higgs number would be quite large 
except the models without Higgs fields.
Therefore, for phenomenological applications, 
it would be important to make the flavor structure richer.
That is the purpose of the next section including 
model building with smaller number of Higgs fields.

\section{Magnetized orbifold models}

\subsection{$U(1)$ gauge theory on magnetized orbifold $T^2/Z_2$}

Now, let us study $U(1)$ gauge theory on the orbifold $T^2/Z_2$ 
with the coordinates $(y_4,y_5)$, 
which are transformed as 
\begin{equation}
y_4 \rightarrow -y_4, \qquad y_5 \rightarrow -y_5,
\end{equation}
under the $Z_2$ orbifold twist.
Then, we introduce the same magnetic flux $F_{45}=2 \pi M$ as 
one in section 2.2 and use the same gauge as (\ref{eq:gauge}).
Note that this magnetic flux is invariant under the $Z_2$ orbifold
twist.

We study the spinor field $\psi(y)$ on the above background.
The spinor field $\psi(y)$ with the $U(1)$ charge $q=\pm 1$ 
satisfies the same equation as one on $T^2$, i.e. (\ref{eq:Dirac-T2}).
Then, we require $\psi(y)$ transform under the $Z_2$ twist as 
\begin{equation}
\psi(-y_4,-y_5) = (-i)\tilde \Gamma^4 \tilde \Gamma^5
P\psi(-y_4,-y_5),
\end{equation}
where $P$ depends on the charge $q$ like $P=(-1)^{q+n}$ 
with $n=$ integer and it should satisfy $P^2 =1$.
Suppose that $qM >0$.
Then, there are $M$ independent zero-modes for 
$\psi$ when we do not take into account the $Z_2$ projection.
However, some of them are projected out 
by the above $Z_2$ boundary condition.
For example, for $(-i)\tilde \Gamma^4 \tilde \Gamma^5P=1$, 
only even functions remain, while 
only odd functions remain for $(-i)\tilde \Gamma^4 \tilde \Gamma^5P=-1$.
Note that 
\begin{equation}
\Theta^j(-y_4,-y_5)=\Theta^{M-j}(y_4,y_5),
\end{equation}
where $\Theta^{M}(y_4,y_5)=\Theta^{0}(y_4,y_5)$.
That is, even and odd functions are given by 
\begin{eqnarray}
\Theta^j_{\rm even}&=& \frac{1}{\sqrt 2}(\Theta^j + \Theta^{M-j}), \\
\Theta^j_{\rm odd}&=& \frac{1}{\sqrt 2}(\Theta^j - \Theta^{M-j}), 
\end{eqnarray}
respectively.
Hence, for $M=2k$ with $k=$ integer and $k>0$, 
the number of zero-modes $\psi_+$ for $P=1$ 
and $P=-1$ are equal to $k+1$ and $k-1$, respectively.
On the other hand, for $M=2k+1$ with $k=$ integer and $k\geq 0$, 
the number of zero-modes $\psi_+$ for $P=1$ 
and $P=-1$ are equal to $k+1$ and $k$, respectively.
It is interesting that odd functions can correspond to zero-modes 
in magnetized orbifold models.
On the orbifold with vanishing magnetic flux $M=0$, 
odd modes correspond to not zero-modes, but massive modes.
However, odd modes, which would correspond to massive modes 
for $M=0$, mix to lead to zero-modes in the case with $M\neq 0$.
It would be convenient to write these results explicitly 
for later discussions.
Table 1 shows the numbers of zero-modes with even 
and odd wave functions for $M \leq 10$.
Note that the degree of constant shift $y_m \rightarrow y_m +d_m$, 
which we have on the torus,  
is ruled out on the orbifold.

\begin{table}[htb]
\begin{center}
\begin{tabular}{|c|ccccccccccc|}\hline
$M$ & 0 & 1 & 2 & 3 & 4& 5 & 6 & 7 & 8 & 9& 10 \\ \hline 
even & 1 & 1 & 2 & 2 & 3 & 3 & 4 & 4 & 5 & 5 & 6 \\ \hline
odd & 0 & 0 & 0& 1 & 1 & 2 & 2 & 3 & 3 & 4 & 4 \\ \hline
\end{tabular}
\end{center}
\caption{The numbers of zero-modes with even and odd wave functions.
\label{even-odd-zero-modes}}
\end{table}

\subsection{$U(N)$ gauge theory on magnetized orbifold $T^2/Z_2$}

Now, let us study $U(N)$ gauge theory on the orbifold $T^2/Z_2$.
We consider the same magnetic flux as (\ref{eq:F45-UN}), 
which breaks the gauge group 
$U(N) \rightarrow \prod_{a=1}^n U(N_a)$.
Furthermore, we associate the $Z_2$ twist with the $Z_2$ 
action in the gauge space as 
\begin{equation}
A_\mu(x,-y) = P A_\mu(x,y)P^{-1}, \qquad A_m(x,y) = -P A_m (x,y)P^{-1}.
\end{equation}
In general, the $Z_2$ boundary condition breaks the gauge group 
$\prod_{a=1}^n U(N_a)$ further.
For simplicity, here we restrict ourselves to 
the $Z_2$ action, which remains the gauge group $\prod_{a=1}^n U(N_a)$ 
unbroken.
Thus, the $Z_2$ action is trivial for the unbroken gauge group, 
but it is not trivial for spinor fields as well as scalar fields.

Here, let us study spinor fields.
We focus on the $U(N_a) \times U(N_b)$ block (\ref{eq:F-ab-block}) 
and use the same gauge as (\ref{eq:gauge}), 
i.e. $A_4 = 0$ and $A_5 = F_{45}y_4$.
We consider the spinor fields, $\lambda^{aa}_\pm$, 
$\lambda^{ab}_\pm$, $\lambda^{ba}_\pm$ and $\lambda^{bb}_\pm$, 
where $\pm$ denotes the chirality in the extra dimension like 
(\ref{eq:two-spinor}).
Their $Z_2$ boundary conditions are given by 
\begin{equation}
\lambda_{\pm}(x,-y) = \pm P \lambda_{\pm}(x,y) P^{-1},
\end{equation}
for $\lambda^{aa}_\pm$, 
$\lambda^{ab}_\pm$, $\lambda^{ba}_\pm$ and $\lambda^{bb}_\pm$.
First, we study the gaugino fields,  $\lambda^{aa}_\pm$ 
and $\lambda^{bb}_\pm$ for the unbroken gauge group.
Since the $Z_2$ action $P$ is trivial for the unbroken gauge 
indices, the above   $Z_2$ boundary conditions reduce to 
$\lambda^{aa}_{\pm}(x,-y)=\pm \lambda^{aa}_{\pm}(x,y)$ and 
$\lambda^{bb}_{\pm}(x,-y)=\pm \lambda^{bb}_{\pm}(x,y)$.
In addition, the magnetic flux does not appear in their 
zero-mode equations.
Thus, $\lambda^{aa}_{+}(x,y)$ as well as $\lambda^{bb}_{+}(x,y)$
has a zero-mode, but $\lambda^{aa}_{-}(x,y)$ and 
$\lambda^{bb}_{-}(x,y)$ are projected out by the $Z_2$ orbifold 
projection as the usual $Z_2$ orbifold without the magnetic flux.

Next, let us study the bi-fundamental matter fields 
$\lambda^{ab}_\pm$ and $\lambda^{ba}_\pm$.
The magnetic flux $M_a - M_b$ appears in their zero-mode 
equations.
Without the $Z_2$ projection, there are $|M_a - M_b|$ 
zero modes.
For example, when $M_a - M_b>0$, $\lambda^{ab}_+$ as well as 
$\lambda^{ba}_-$ has  $(M_a - M_b)$ zero modes with 
the wave functions $\Theta^j$ for $j=0, \cdots, (M_a - M_b -1)$.
When we consider the $Z_2$ projection, either even or odd modes 
remain.
For example, when we consider the projection $P$ such that 
$\lambda^{ab}_+(x,-y) =\lambda^{ab}_+(x,y)$,  
only zero-modes corresponding to $\Theta^j_{\rm even}$ remain 
and the number of zero-modes is equal to $(M_a - M_b)/2 + 1$ for 
$(M_a - M_b)=$ even and $(M_a - M_b+1)/2 $ for $(M_a - M_b)=$ odd.
On the other hand, when we consider the projection $P$ such that 
$\lambda^{ab}_+(x,-y) =-\lambda^{ab}_+(x,y)$,  
only zero-modes corresponding to $\Theta^j_{\rm odd}$ remain 
and the number of zero-modes is equal to $(M_a - M_b)/2 - 1$ for 
$(M_a - M_b)=$ even and $(M_a - M_b - 1)/2 $ for $(M_a - M_b)=$ odd.
The same holds true for $\lambda^{ba}_-$.
Furthermore, when $M_a - M_b <0$, the situation is the same 
except replacing  $(M_a - M_b)$, $\lambda^{ab}_+$ and
$\lambda^{ba}_-$ by $|M_a - M_b|$, $\lambda^{ab}_-$ and
$\lambda^{ba}_+$, respectively.

The 3-point couplings among modes corresponding to 
the wave functions, $\Theta^i_{\rm even,odd}$, 
$\Theta^j_{\rm even,odd}$ and $\Theta^k_{\rm even,odd}$ are 
given by the overlap integral like (\ref{eq:Y-overlap}).
Note that 
\begin{equation}\label{eq:vanish-overlap}
\int dy~\Theta^i_{\rm even}(y) \cdot \Theta^j_{\rm even}(y) 
\cdot \Theta^k_{\rm odd}(y) = 
\int dy~\Theta^i_{\rm odd}(y) \cdot \Theta^j_{\rm odd}(y) 
\cdot \Theta^k_{\rm odd}(y) = 0,
\end{equation}
while $\int dy~\Theta^i_{\rm even}(y) \cdot \Theta^j_{\rm odd}(y) 
\cdot \Theta^k_{\rm odd}(y) $ and 
$\int dy~\Theta^i_{\rm even}(y) \cdot \Theta^j_{\rm even}(y) 
\cdot \Theta^k_{\rm even}(y) $ are nonvanishing.

\subsection{$U(N)$ gauge theory on magnetized orbifolds 
$T^6/Z_2$ and $T^6/(Z_2\times Z'_2)$}

Here, we can extend the previous analysis on 
the two-dimensional orbifold $T^2/Z_2$ to 
the $U(N)$ gauge theory on the six-dimensional 
orbifolds $T^6/Z_2$ and $T^6/(Z_2\times Z'_2)$.
We consider two types of six-dimensional orbifolds, 
$T^6/Z_2$ and $T^6/(Z_2\times Z'_2)$.
For the orbifold $T^6/Z_2$, the $Z_2$ twist 
acts on the six-dimensional coordinates $y_m$ ($m=4,\cdots, 9$) as 
\begin{equation}
y_m \rightarrow -y_m~~({\rm for~~}m=4,5,6,7), \qquad 
y_n \rightarrow y_n~~({\rm for~~}n=8,9).
\end{equation}
In addition to this $Z_2$ action, we introduce another 
independent $Z'_2$ action,
\begin{equation}
y_m \rightarrow -y_m~~({\rm for~~}m=4,5,8,9), \qquad 
y_n \rightarrow y_n~~({\rm for~~}n=6,7),
\end{equation}
for the orbifold $T^6/(Z_2\times Z'_2)$.
If magnetic flux is vanishing, we realize 
four-dimensional N=2 and N=1 supersymmetric gauge 
theories for the orbifolds,  $T^6/Z_2$ and 
$T^6/(Z_2\times Z'_2)$, respectively.

Now, let us introduce 
the same magnetic flux as (\ref{eq:6D-flux}).
The gauge group $U(N)$ is broken as 
$U(N) \rightarrow \prod_{a=1}^n U(N_a)$ with $N=\sum_a N_a$.
This magnetic flux is invariant under 
both $Z_2$ and $Z'_2$ actions.
Furthermore, we associate the $Z_2$ and $Z'_2$ twists with 
the $Z_2$ and $Z'_2$ actions in the gauge space as 
\begin{eqnarray}
A_\mu(x,-y_m,y_n)   &=& PA_\mu(x,-y_m,y_n)P^{-1}, \nonumber \\
A_m(x,-y_m,y_n)   &=& -PA_m(x,-y_m,y_n)P^{-1}, \\
A_n(x,-y_m,y_n)   &=& PA_n(x,-y_m,y_n)P^{-1}, \nonumber
\end{eqnarray}
for $m=4,5,6,7$ and $n=8,9$, and 
\begin{eqnarray}
A_\mu(x,-y_m,y_n)   &=& P'A_\mu(x,-y_m,y_n){P'}^{-1}, \nonumber \\
A_m(x,-y_m,y_n)   &=& -P'A_m(x,-y_m,y_n){P'}^{-1}, \\
A_n(x,-y_m,y_n)   &=& P'A_n(x,-y_m,y_n){P'}^{-1}, \nonumber
\end{eqnarray}
for $m=4,5,8,9$ and $n=6,7$.
In general, these $Z_2$ boundary conditions break the gauge group 
$\prod_{a=1}^n U(N_a)$ further.
For simplicity, here we restrict to the $Z_2$ and $Z_2'$ 
projections, which remain the gauge group $\prod_{a=1}^n U(N_a)$ 
unbroken.
That is, both the $Z_2$ and $Z_2'$ actions are trivial 
for the unbroken gauge group.

Now, we study spinor fields.
We focus on the $U(N_a) \times U(N_b)$ block as
(\ref{eq:6D-F-ab-block}) and use the same gauge as
(\ref{eq:6D-gauge}).
We consider the spinor fields $\lambda^{aa}_{s_1,s_2,s_3}$, 
$\lambda^{ab}_{s_1,s_2,s_3}$, $\lambda^{ba}_{s_1,s_2,s_3}$ 
and $\lambda^{bb}_{s_1,s_2,s_3}$, where $s_i$ 
denotes the chirality corresponding to the $i$-th $T^2$.
Their $Z_2$ boundary conditions are given by 
\begin{equation}
\lambda_{s_1,s_2,s_3}(x,-y_m,y_n) =
s_1s_2P\lambda_{s_1,s_2,s_3}(x,y_m,y_n) P^{-1},
 \end{equation}
with $m=4,5,6,7$ and $n=8,9$ for $\lambda^{aa}_{s_1,s_2,s_3}$, 
$\lambda^{ab}_{s_1,s_2,s_3}$, $\lambda^{ba}_{s_1,s_2,s_3}$ 
and $\lambda^{bb}_{s_1,s_2,s_3}$.
Similarly, the $Z'_2$ boundary conditions are given by 
\begin{equation}
\lambda_{s_1,s_2,s_3}(x,-y_m,y_n) =
s_1s_3P'\lambda_{s_1,s_2,s_3}(x,y_m,y_n) P'^{-1},
 \end{equation}
with $m=4,5,8,9$ and $n=6,7$.

First, we study the gaugino fields $\lambda^{aa}_{s_1,s_2,s_3}$ 
and $\lambda^{bb}_{s_1,s_2,s_3}$ for the unbroken gauge group.
Their zero-mode equations have no effect due to 
magnetic fluxes, but only the $Z_2$ and $Z'_2$ orbifold twists 
play a role.
Since the $Z_2$ and $Z'_2$ twists, $P$ and $P'$, are trivial 
for the unbroken gauge sector, the boundary conditions are 
given by 
\begin{equation}
\lambda^{aa(bb)}_{s_1,s_2,s_3}(x,-y_m,y_n) =
s_1s_2\lambda^{aa(bb)}_{s_1,s_2,s_3}(x,y_m,y_n) \qquad{\rm for~~} Z_2,
 \end{equation}
with $m=4,5,6,7$ and $n=8,9$, and 
\begin{equation}
\lambda^{aa(bb)}_{s_1,s_2,s_3}(x,-y_m,y_n) =
s_1s_3\lambda^{aa(bb)}_{s_1,s_2,s_3}(x,y_m,y_n) \qquad{\rm for~~} Z'_2,
 \end{equation}
with $m=4,5,8,9$ and $n=6,7$.
Hence, zero modes of $\lambda^{aa(bb)}_{+,+,\pm}$ and 
$\lambda^{aa(bb)}_{-,-,\pm}$ survive on $T^6/Z_2$, 
that is, two kinds of gaugino fields with a fixed 
four-dimensional chirality.
Furthermore, on $T^6/(Z_2 \times Z'_2)$, 
zero modes of $\lambda^{aa(bb)}_{+,+,+}$ and 
$\lambda^{aa(bb)}_{-,-,-}$ survive, that is, a single sort of 
gaugino fields with a fixed four-dimensional chirality.

Next, let us study the bi-fundamental matter fields, 
$\lambda^{ab}_{s_1,s_2,s_3}$ and $\lambda^{ba}_{s_1,s_2,s_3}$.
Without the $Z_2$ projection, they have zero-modes, 
whose number is $I_{ab}=I^1_{ab}I^{2}_{ab}I^3_{ab}$ and 
wave functions are given by 
$\Theta^{j_1}(y_4,y_5) \Theta^{j_2}(y_6,y_7) \Theta^{j_3}(y_8,y_9)$ 
($j_i=0,\cdots,(I^i_{ab}-1)$).
We assume that $I^i_{ab} > 0$ for $i=1,2,3$.
Then, the zero-modes correspond to $\lambda^{ab}_{+,+,+}$.
On $T^6/Z_2$, some of them are projected out.
Suppose that the $Z_2$ boundary condition is given by 
\begin{equation}
\lambda^{ab}_{s_1,s_2,s_3}(x,-y_m,y_n) =
s_1s_2\lambda^{ab}_{s_1,s_2,s_3}(x,y_m,y_n),
 \end{equation}
with $m=4,5,6,7$ and $n=8,9$.
Then, surviving zero-modes correspond to 
$\Theta^{j_1}_{\rm even}(y_4,y_5) \Theta^{j_2}_{\rm even}(y_6,y_7) 
\Theta^{j_3}(y_8,y_9)$ and 
$\Theta^{j_1}_{\rm odd}(y_4,y_5) \Theta^{j_2}_{\rm odd}(y_6,y_7) 
\Theta^{j_3}(y_8,y_9)$.
Further modes are projected out on $T^6/(Z_2 \times Z'_2)$.
Suppose that the $Z'_2$ boundary condition is given by 
\begin{equation}
\lambda^{ab}_{s_1,s_2,s_3}(x,-y_m,y_n) =
s_1s_3\lambda^{ab}_{s_1,s_2,s_3}(x,y_m,y_n),
 \end{equation}
with $m=4,5,8,9$ and $n=6,7$.
Then, the surviving modes through the $Z_2 \times Z'_2$ projection 
correspond to 
$\Theta^{j_1}_{\rm even}(y_4,y_5) \Theta^{j_2}_{\rm even}(y_6,y_7) 
\Theta^{j_3}_{\rm even}(y_8,y_9)$ and 
$\Theta^{j_1}_{\rm odd}(y_4,y_5) \Theta^{j_2}_{\rm odd}(y_6,y_7) 
\Theta^{j_3}_{\rm odd}(y_8,y_9)$.
Similarly, we can analyze surviving zero-modes through 
the $Z_2 \times Z'_2$ projection in the models with 
different signs of $I^i_{ab}$ and different $Z_2 \times Z'_2$ 
projections.
It would be convenient to introduce the notation, 
$I^i_{ab({\rm even})}$ and $I^i_{ab({\rm odd})}$, 
such that  $I^i_{ab({\rm even})}$ and $I^i_{ab({\rm odd})}$ 
denote the number of even and odd functions, $\Theta^j_{\rm even}$ 
and $\Theta^j_{\rm odd}$, respectively, among $|I^i_{ab}|$ functions 
$\Theta^j$ for the $i$-th $T^2$.
Note that $I^i_{ab({\rm even})}, I^i_{ab({\rm odd})} \geq 0$ 
in the above definition, while $I^i_{ab}$ can be negative.

\subsection{$U(N_a) \times U(N_b) \times U(N_c)$ models 
with three families} 

Here, we consider the $U(N_a) \times U(N_b) \times U(N_c)$ models, 
which are derived from ten-dimensional $U(N)$ super Yang-Mills theory 
on $R^{3,1}\times T^6/(Z_2 \times Z'_2)$ with the same magnetic flux 
as in the previous subsection, e.g. $N_a=4, N_b=2$ and $N_c=2$.
Suppose that four-dimensional N=1 supersymmetry remains.
In addition to the $U(N_a) \times U(N_b) \times U(N_c)$ 
vector multiplets, the massless spectrum includes 
three types of bi-fundamental fields, 
$(N_a,\bar N_b)$, $(N_b,\bar N_c)$ and $(N_c,\bar N_a)$.
As section 2.5, we assign  $(N_a,\bar N_b)$ and $(\bar N_a,N_c)$ 
to left-handed and right-handed matter fields.
In this case, $(N_b,\bar N_c)$ modes correspond to 
Higgs fields.

%To summarize the recipe of model building, 
Now, we give explicit models.
For simplicity, we restrict ourselves to models, which satisfy 
the condition (\ref{eq:SUSY-condition}).
%We start with the ten-dimensional $U(N)$ super Yang-Mills 
%theory on the background $R^{3,1}\times T^6/(Z_2 \times Z'_2)$.
For example, we introduce the following magnetic flux,
\begin{eqnarray}\label{eq:model-1}
F_{45} &=& 2 \pi\left(
\begin{array}{ccc}
0 \times {\bf 1}_{N_a \times N_a} & & 0  \\
 & -3 \times {\bf 1}_{N_b \times N_b} & \\
0 & & -4 \times {\bf 1}_{N_c\times N_c}
\end{array}\right),  \nonumber \\
F_{67} &=& 2 \pi \left(
\begin{array}{ccc}
0 \times  {\bf 1}_{N_a \times N_a} & & 0 \\
 & -4 \times {\bf 1}_{N_b \times N_b} & \\
0 & & -1 \times {\bf 1}_{N_c\times N_c}
\end{array}\right),  \\
F_{89} &=& 2 \pi \left(
\begin{array}{ccc}
0 \times  {\bf 1}_{N_a \times N_a} & & 0 \\
 & -1 \times {\bf 1}_{N_b \times N_b} & \\
0 & & 3 \times {\bf 1}_{N_c\times N_c}
\end{array}\right). \nonumber 
\end{eqnarray}
This magnetic flux breaks the gauge group 
$U(N) \rightarrow U(N_a)\times U(N_b) \times U(N_c)$, 
and satisfies the condition (\ref{eq:SUSY-condition}).
In addition, we consider the orbifold projections, e.g. 
\begin{equation}\label{eq:z2-P}
P = P' = \left( 
\begin{array}{ccc}
{\bf 1}_{N_a \times N_a} & & 0  \\
 & {\bf -1}_{N_b \times N_b} & \\
0 & & {\bf 1}_{N_c\times N_c}
\end{array}\right),
\end{equation}
which do not break $U(N_a)\times U(N_b) \times U(N_c)$.
A single sort of $U(N_a)\times U(N_b) \times U(N_c)$ gaugino fields 
remain through the orbifold projection.

Now, let us study the spinor fields $\lambda^{ab}$.
Their Dirac equations have the difference of magnetic fluxes, 
$I^i_{ab}=(3,4,1)$.
Thus, their zero-modes correspond to $\lambda^{ab}_{+,+,+}$, 
which transform 
$\lambda^{ab}_{+,+,+}(x,-y_m,y_n) \rightarrow - 
\lambda^{ab}_{+,+,+}(x,y_m,y_n)$
for both $Z_2$ and $Z'_2$ actions.
In general, these boundary conditions are satisfied with 
both types of the wave functions 
$\Theta^{j_1}_{\rm odd}(y_4,y_5)\Theta^{j_2}_{\rm even}(y_6,y_7)
\Theta^{j_3}_{\rm even}(y_8,y_9)$ and 
$\Theta^{j_1}_{\rm even}(y_4,y_5)\Theta^{j_2}_{\rm odd}(y_6,y_7)
\Theta^{j_3}_{\rm odd}(y_8,y_9)$.
However, note that this model has $I^3_{ab}=1$ and 
$I^3_{ab({\rm odd})}=0$, that is, there is no zero-mode 
corresponding to $\Theta^{j_3}_{\rm odd}(y_8,y_9)$.
Thus, the zero-modes correspond to only the wave functions 
$\Theta^{j_1}_{\rm odd}(y_4,y_5)\Theta^{j_2}_{\rm even}(y_6,y_7)
\Theta^{j_3}_{\rm even}(y_8,y_9)$ and 
the total number of zero-modes is given by 
the product of 
$I^1_{ab({\rm odd})}=1$, $I^2_{ab({\rm even})}=3$ and 
$I^3_{ab({\rm even})}=1$, that is, there are three 
zero-modes.
The magnetic flux difference, $I^i_{ab}=(3,4,1)$,
%is the same as one in the model (\ref{eq:12-f-model}), 
which would have twelve families of $(N_a,\bar N_b)$ without 
orbifolding.
However, the orbifold projection reduces the family 
number from twelve to three.
Similarly, we can analyze zero-modes for  $\lambda^{bc}$ and 
$\lambda^{ca}$.  
The result is shown in Table 2.
The second column shows magnetic fluxes, which appear in 
their Dirac equations, and the subscript $ef$ denotes 
$ef=ab, ca$ and $bc$.
The third and fourth columns show six-dimensional chiralities 
of zero-modes and their forms of wave functions.
The fifth column shows the total number of zero-modes.
This model has three families when we consider 
$\lambda^{ab}$ and $\lambda^{ca}$ as left-handed and right-handed 
matter fields.
The scalar fields associated with $\lambda^{bc}$ would correspond 
to Higgs fields.
However, their Yukawa couplings are not allowed in this model, 
because of the periodicity (\ref{eq:vanish-overlap}).

\begin{table}[htb]
\begin{center}
\begin{tabular}{|c|c|c|c|c|} \hline
     & $I^i_{ef}$ & chirality & wave function & the total number  \\
     &            &           &               & of zero-modes \\ \hline 
$\lambda^{ab}$ & $(3,4,1)$ & $\lambda^{ab}_{+,+,+}$ & 
$\Theta^{j_1}_{\rm odd}\Theta^{j_2}_{\rm even}
\Theta^{j_3}_{\rm even}$ & 3 \\ 
$\lambda^{ca}$ & $(-4,-1,3)$ & $\lambda^{ca}_{-,-,+}$ & 
$\Theta^{j_1}_{\rm even}\Theta^{j_2}_{\rm even}
\Theta^{j_3}_{\rm odd}$ & 3 \\ 
$\lambda^{bc}$ & $(1,-3,-4)$ & $\lambda^{bc}_{+,-,-}$ & 
$\Theta^{j_1}_{\rm even}\Theta^{j_2}_{\rm even}
\Theta^{j_3}_{\rm even}$ & 6 \\ 
\hline \end{tabular}
\end{center}
\caption{Three-family model.
\label{Model}}
\end{table}

We show another model with the following magnetic flux,
\begin{eqnarray}\label{eq:model-2}
F_{45} &=& 2 \pi \left(
\begin{array}{ccc}
0 \times {\bf 1}_{N_a \times N_a} & & 0  \\
 & -3 \times {\bf 1}_{N_b \times N_b} & \\
0 & &  {\bf 1}_{N_c\times N_c}
\end{array}\right),  \nonumber \\
F_{67} &=& 2 \pi \left(
\begin{array}{ccc}  
0 \times  {\bf 1}_{N_a \times N_a} & & 0 \\
 & -4 \times {\bf 1}_{N_b \times N_b} & \\
0 & & -5 \times  {\bf 1}_{N_c\times N_c}
\end{array}\right),  \\
F_{89} &=& 2 \pi \left(
\begin{array}{ccc}  
  0 \times {\bf 1}_{N_a \times N_a} & & 0 \\
 & -1 \times {\bf 1}_{N_b \times N_b} & \\
0 & & -4 \times {\bf 1}_{N_c\times N_c}
\end{array}\right). \nonumber 
\end{eqnarray}
This magnetic flux breaks the gauge group 
$U(N) \rightarrow U(N_a)\times U(N_b) \times U(N_c)$, 
and satisfies the condition (\ref{eq:SUSY-condition}).
We consider the following orbifold projections,
\begin{eqnarray}
P &=& \left( 
\begin{array}{ccc}
{\bf -1}_{N_a \times N_a} & & 0  \\
 & {\bf 1}_{N_b \times N_b} & \\
0 & & {\bf 1}_{N_c\times N_c}
\end{array}\right), \nonumber \\
P' &=& \left( 
\begin{array}{ccc}
{\bf 1}_{N_a \times N_a} & & 0  \\
 & {\bf -1}_{N_b \times N_b} & \\
0 & & {\bf 1}_{N_c\times N_c}
\end{array}\right).
\end{eqnarray}%We use the same orbifold projections as (\ref{eq:z2-P}).
Then, we can analyze the zero-modes as the above.
The result for bi-fundamental matter is shown in Table 3.
This model has three families of $\lambda^{ab}$ and $\lambda^{ca}$.
The scalar fields associated with $\lambda^{bc}$ can couple with 
them.
The three families of $\lambda^{ab}$ and $\lambda^{ca}$ are 
quasi-localized at points different from each other on the second $T^2$.
Furthermore, one family of $\lambda^{ab}$ and $\lambda^{ca}$ 
is quasi-localized at the same point as the point, 
where  $\lambda^{bc}$ is quasi-localized.
That could explain one family has a large Yukawa coupling with 
the Higgs fields, while the other families have smaller 
Yukawa couplings.
However, the full form of Yukawa matrices is not 
completely realistic, because the up and down sectors of 
quarks have the same form of Yukawa matrices.
We would study Yukawa matrices numerically elsewhere 
taking into account $SU(2)_R$ breaking.

Similarly, we can construct other three-family 
models, which satisfy the condition (\ref{eq:SUSY-condition}).
Also the model construction can be done in a similar way
when we do not take into account the condition (\ref{eq:SUSY-condition}).

\begin{table}[htb]
\begin{center}
\begin{tabular}{|c|c|c|c|c|} \hline
     & $I^i_{ef}$ & chirality & wave function & the total number  \\
     &            &           &               & of zero-modes \\ \hline 
$\lambda^{ab}$ & $(3,4,1)$ & $\lambda^{ab}_{+,+,+}$ & 
$\Theta^{j_1}_{\rm odd}\Theta^{j_2}_{\rm even}
\Theta^{j_3}_{\rm even}$ & 3 \\ 
$\lambda^{ca}$ & $(1,-5,-4)$ & $\lambda^{ca}_{+,-,-}$ & 
$\Theta^{j_1}_{\rm even}\Theta^{j_2}_{\rm even}
\Theta^{j_3}_{\rm odd}$ & 3 \\ 
$\lambda^{cb}$ & $(4,-1,-3)$ & $\lambda^{cb}_{+,-,-}$ & 
$\Theta^{j_1}_{\rm odd}\Theta^{j_2}_{\rm even}
\Theta^{j_3}_{\rm odd}$ & 1 \\ 
\hline \end{tabular}
\end{center}
\caption{Three-family model.
\label{Model-2}}
\end{table}

These two models are not completely realistic, but 
simple models to illustrate explicit model building.
One of important features is that the family number is 
smaller than the magnetized torus models with the same 
magnetic fluxes and there are a variety of 
models with a fixed number of families, e.g. 
three-family models.
Another important feature in generic model is that 
adjoint matter fields except gaugino fields are 
projected out by the orbifold projection on $T^6/(Z_2 \times Z'_2)$, 
although they remain on $T^6/Z_2$.
Its implication from the viewpoint of model building 
is as follows.
The above two models would correspond to 
the three-family Pati-Salam model when 
$N_a=4$, $N_b=2$ and $N_c=2$.
In intersecting D-brane models, the Pati-Salam gauge 
group is broken by splitting D-branes to realize
the breaking $U(4) \rightarrow U(3) \times U(1)$ 
and $U(2) \rightarrow U(1)^2$ and 
such splitting correspond to the symmetry breaking 
by vacuum expectation values of adjoint scalar fields.
However, we have no degree of freedom of adjoint scalar fields.
One of simple ways to realize the standard gauge group 
in the above model building is that we start with 
$U(6)\times U(1)'_a \times U(1)'_c$.
We introduce the same magnetic fluxes as (\ref{eq:model-1}) and 
(\ref{eq:model-2}) for $N_a=3$, $N_b=2$ and $N_c=1$ in 
$U(6)=U(N_a+N_b+N_c)$.
In addition, we introduce the same magnitude of magnetic fluxes in 
$ U(1)'_a$ and $U(1)'_c$ as 
those in $U(N_a)$ and  $U(N_c)$ blocks of 
$U(N)$, respectively.
Then we can obtain supersymmetic standard models with 
three families of quarks and leptons.

Alternatively, we have the degree of freedom to 
introduce any field on the orbifold fixed points.
That is, we can break the Pati-Salam gauge group 
into the standard gauge group by vacuum 
expectation values of brane modes like 
the adjoint scalar field or a pair of the 
Higgs fields $(4,1,2)$ and $(\bar 4,1,2)$.
Another way to break the gauge symmetry is to 
change the orbifold projection such that 
$P$ and $P'$ break the gauge group further.

In addition to the above Higgs fields, one can 
introduce any mode on the orbifold fixed points.
For example, all of three families may not be originated from 
bulk modes, but some of quarks and leptons are originated from 
such brane modes.
That is, we have an interesting variety for model building.
In Appendix, we give two examples of models, whose 
family numbers of bulk modes are not equal to three.
Furthermore, such brane modes can not couple with 
bulk modes, whose wave functions include 
$\Theta^j_{\rm odd}$ for the $i$-th $T^2$, 
because  the wave function $\Theta^j_{\rm odd}$ vanishes 
on the fixed point.
That is a new aspect in our model.
In the usual orbifold models, bulk zero-modes correspond to 
even functions.
Thus, they can couple with brane modes.
However, in our model some of bulk modes can not couple with 
brane modes.
This fact would be important for further model building.

\section{Conclusion}

We have studied D-dimensional N=1 super 
Yang-Mills theory on the orbifold background 
with non-vanishing magnetic fluxes, 
in particular spinor fields.
Our models have a rich structure.
Odd modes can have zero-modes 
and couplings are controlled by 
the orbifold periodicity of wave functions.
We can derive flavor structures different from 
those in magnetized torus models.
Thus, further study on model building 
would be interesting.

We have shown rather simple model building,
although we could consider more generic class
of magnetized orbifold models.
We have more degrees of freedom of extensions for 
model building.
One extension is to introduce brane modes as 
mentioned in section 3.4.
In addition, we can introduce magnetic fluxes 
on orbifold fixed points, which would be
independent of bulk magnetic fluxes and/or 
magnetic fluxes on different fixed points.\footnote{
See e.g. \cite{Lee:2003mc}.}

We have started with D-dimensional N=1 super Yang-Mills theory.
However, we can add hypermultiplets e.g. for D=6.
Also we have considered six-dimensions and ten-dimensions, 
but we can consider eight-dimensional theory in a similar way.
Moreover, we can extend our analysis to 
several combinations of branes, whose dimensions 
are different from each other like 
D9-D5 branes and other combinations.

We have restricted abelian magnetic fluxes, 
but in general non-Abelian magnetic fluxes are 
possible, i.e. the toron background.
That reduces the rank of the gauge group.
Furthermore, we can choose the orbifold projections, 
which break the gauge group further.
Moreover, we have considered the factorizable orbifold 
and non-vanishing magnetic fluxes $F_{2m, 2m+1}$ for $m=2,3,4$.
We could extend to non-factorizable orbifolds \cite{Forste:2006wq} 
and more generic form of magnetic fluxes.

Thus, we have various directions of extensions in 
generic magnetized orbifold models.
Including these extensions, the structure of models 
would become much richer.
Hence, further studies with these extensions are 
quite important.

\subsection*{Acknowledgement}
H.~A.\/ and T.~K.\/ are supported in part by the
Grant-in-Aid for Scientific Research No.~182496 
and No.~20540266 from the Ministry of Education, Culture,
Sports, Science and Technology of Japan.

\appendix

\section{Models}

Here we give two examples of models, whose family numbers 
of bulk modes differ from three.
That is, one model has two bulk families and 
the other has eighteen bulk families.
We start with the ten-dimensional $U(N)$ super Yang-Mills theory 
on the background $R^{3,1}\times T^6/(Z_2 \times Z'_2)$.
We consider the trivial orbifold projections 
$P=P'=1$.

In the first model, we introduce the following magnetic flux,
\begin{eqnarray}\label{eq:model-3}
F_{45} &=& \left(
\begin{array}{ccc}
0 \times {\bf 1}_{N_a \times N_a} & & 0  \\
 & -2 \times {\bf 1}_{N_b \times N_b} & \\
0 & & 2 \times {\bf 1}_{N_c\times N_c}
\end{array}\right),  \nonumber \\
F_{67} &=& \left(
\begin{array}{ccc}  
0 \times  {\bf 1}_{N_a \times N_a} & & 0 \\
 & -1 \times {\bf 1}_{N_b \times N_b} & \\
0 & & 1 \times  {\bf 1}_{N_c\times N_c}
\end{array}\right),  \\
F_{89} &=& \left(
\begin{array}{ccc}  
  0 \times {\bf 1}_{N_a \times N_a} & & 0 \\
 & -1 \times {\bf 1}_{N_b \times N_b} & \\
0 & & 1 \times {\bf 1}_{N_c\times N_c}
\end{array}\right). \nonumber 
\end{eqnarray}
This magnetic flux satisfies the condition (\ref{eq:SUSY-condition}) 
and breaks the gauge group 
$U(N) \rightarrow U(N_a)\times U(N_b) \times U(N_c)$, 
although the orbifold projections are trivial $P=P'=1$.
Then, we can analyze the zero-modes as section 3.4.
The result is shown in Table 4.
This model has two bulk families, when we consider $\lambda^{ab}$ 
and $\lambda^{ca}$ as left-handed and right-handed matter fields.
This flavor number is not realistic.
%, and 
%corresponds to the flavor number of the model 
%in section 2.5 (\ref{eq:two-f-model}). 
However, in orbifold models 
it is possible to assume that one family appears on one of fixed points.

\begin{table}[htb]
\begin{center}
\begin{tabular}{|c|c|c|c|c|} \hline
     & $I^i_{ef}$ & chirality & wave function & the total number  \\
     &            &           &               & of zero-modes \\ \hline 
$\lambda^{ab}$ & $(2,1,1)$ & $\lambda^{ab}_{+,+,+}$ & 
$\Theta^{j_1}_{\rm even}\Theta^{j_2}_{\rm even}
\Theta^{j_3}_{\rm even}$ & 2 \\ 
$\lambda^{ca}$ & $(2,1,1)$ & $\lambda^{ca}_{+,+,+}$ & 
$\Theta^{j_1}_{\rm even}\Theta^{j_2}_{\rm even}
\Theta^{j_3}_{\rm even}$ & 2 \\ 
$\lambda^{cb}$ & $(4,2,2)$ & $\lambda^{cb}_{+,+,+}$ & 
$\Theta^{j_1}_{\rm even}\Theta^{j_2}_{\rm even}
\Theta^{j_3}_{\rm even}$ & 12 \\ 
\hline \end{tabular}
\end{center}
\caption{Two-family model from the bulk.
\label{Model-3}}
\end{table}

We give another example.
We use the same orbifold projections, i.e. $P=P'=1$.
We introduce the following magnetic flux,
\begin{eqnarray}\label{eq:model-4}
F_{45} &=& \left(
\begin{array}{ccc}
0 \times {\bf 1}_{N_a \times N_a} & & 0  \\
 & -6 \times {\bf 1}_{N_b \times N_b} & \\
0 & & 6 \times {\bf 1}_{N_c\times N_c}
\end{array}\right),  \nonumber \\
F_{67} &=& \left(
\begin{array}{ccc}  
0 \times  {\bf 1}_{N_a \times N_a} & & 0 \\
 & -3 \times {\bf 1}_{N_b \times N_b} & \\
0 & & 3 \times  {\bf 1}_{N_c\times N_c}
\end{array}\right),  \\
F_{89} &=& \left(
\begin{array}{ccc}  
  0 \times {\bf 1}_{N_a \times N_a} & & 0 \\
 & -3 \times {\bf 1}_{N_b \times N_b} & \\
0 & & 3 \times {\bf 1}_{N_c\times N_c}
\end{array}\right). \nonumber 
\end{eqnarray}
We study the spinor fields $\lambda^{ab}$, in whose Dirac 
equations the difference of magnetic fluxes 
$I^i_{ab}=(6,3,3)$ appears.
Their zero-modes correspond to $\lambda^{ab}_{+,+,+}$, 
which transform 
$\lambda^{ab}_{+,+,+}(x,y_m,y_n) \rightarrow \lambda^{ab}_{+,+,+}(x,-y_m,y_n)$
for both $Z_2$ and $Z'_2$ actions.
These boundary conditions are satisfied with 
the wave functions  
$\Theta^{j_1}_{\rm even}(y_4,y_5) \Theta^{j_2}_{\rm even}(y_6,y_7) 
\Theta^{j_3}_{\rm even}(y_8,y_9) $ and 
$\Theta^{j_1}_{\rm odd}(y_4,y_5) \Theta^{j_2}_{\rm odd}(y_6,y_7) 
\Theta^{j_3}_{\rm odd}(y_8,y_9) $.
The number of zero-modes corresponding to 
the former wave functions is given by the product of 
$I^1_{ab({\rm even})}=4$, $I^2_{ab({\rm even})}=2$ and 
$I^3_{ab({\rm even})}=2$, while the zero-mode number 
corresponding to the latter is given by the product of 
$I^1_{ab({\rm odd})}=2$, $I^2_{ab({\rm odd})}=1$ and 
$I^3_{ab({\rm odd})}=1$.
Thus, the total number of $\lambda^{ab}$ zero-modes is 
equal to 18$(=16+2)$.
Similarly, we can analyze zero-modes for $\lambda^{bc}$ 
and $\lambda^{ca}$.
The result is shown in Table 5.
For these zero-modes, only two forms of wave functions 
are allowed, that is, one is 
$\Theta^{j_1}_{\rm even}(y_4,y_5) \Theta^{j_2}_{\rm even}(y_6,y_7) 
\Theta^{j_3}_{\rm even}(y_8,y_9) $ and the other is 
$\Theta^{j_1}_{\rm odd}(y_4,y_5) \Theta^{j_2}_{\rm odd}(y_6,y_7) 
\Theta^{j_3}_{\rm odd}(y_8,y_9) $.
The numbers of zero-modes corresponding to the former 
and latter are shown in the third and fourth columns.
Six-dimensional chirality of all zero-modes 
correspond to $\lambda_{+,+,+}$ and they are 
omitted in the table.

This model has 18 families.
It seems that this family number is too large.
However, we can reduce the light family number 
if we assume anti-families of $(\bar N_a,N_b)$ and 
$(N_a,\bar N_c)$ matter fields on fixed points 
and  their mass terms with the above families of matter fields.
Such mass terms are possible for zero-modes 
corresponding to $\Theta^{j_1}_{\rm even}(y_4,y_5) 
\Theta^{j_2}_{\rm even}(y_6,y_7) \Theta^{j_3}_{\rm even}(y_8,y_9) $.
Thus, when we assume $n$ anti-families, the number of 
light families reduces to $(18-n)$.
This type of models has an interesting aspect, that is, 
some families of matter fields correspond to 
$\Theta^{j_1}_{\rm even}(y_4,y_5) 
\Theta^{j_2}_{\rm even}(y_6,y_7) \Theta^{j_3}_{\rm even}(y_8,y_9) $ 
and other families of matter fields correspond to 
$\Theta^{j_1}_{\rm odd}(y_4,y_5) \Theta^{j_2}_{\rm odd}(y_6,y_7) 
\Theta^{j_3}_{\rm odd}(y_8,y_9) $.
In general, other combinations of wave functions can appear 
in zero-modes of matter fields.
Such asymmetry appears in this type of models.
Thus, their flavor structure is rich.

\begin{table}[htb]
\begin{center}
\begin{tabular}{|c|c|c|c|c|} \hline
     & $I^i_{ef}$ & No. of zero-modes & No. of
     zero-modes & the total number  \\
     &            &  $\Theta^{j_1}_{\rm even}\Theta^{j_2}_{\rm even}
\Theta^{j_3}_{\rm even}$  &   $\Theta^{j_1}_{\rm odd}\Theta^{j_2}_{\rm odd}
\Theta^{j_3}_{\rm odd}$         & of zero-modes \\ \hline 
$\lambda^{ab}$ & $(6,3,3)$ & 16 & 2 & 18 \\ 
$\lambda^{ca}$ & $(6,3,3)$ & 16 & 2 & 18 \\
$\lambda^{cb}$ & $(12,6,6)$ & 112 & 20 & 132 \\ 
\hline \end{tabular}
\end{center}
\caption{Eighteen-family model from the bulk.
\label{Model-4}}
\end{table}

\end{document}